\documentclass[aps,prl,superscriptaddress,amsmath,twocolumn,showpacs]{revtex4}
\usepackage{hangcaption,hhline,psfrag,rotating,amssymb}
\usepackage[hang,nooneline]{subfigure}
\usepackage{dcolumn}
\usepackage{bm}
\usepackage[pdftex]{color}
\usepackage[sort&compress]{natbib}
\usepackage[colorlinks=true,linkcolor=blue,filecolor=blue,urlcolor=blue,citecolor=blue,pdftex=true,plainpages=false]{hyperref}
\newcommand{\spose}[1]{\hbox to 0pt{#1\hss}}

\newcommand{\gtapprox}{\gtrsim}
\newcommand{\inapprox}{\mathrel{\spose{\lower 3pt\hbox{$\mathchar"218$}}
 \raise 2.0pt\hbox{$\mathchar"232$}}}

\begin{document}
\author{A.~Bazavov}
\affiliation{Physics Department, Brookhaven National Laboratory, Upton, NY 11973, USA}
\author{C.~Bernard}
\email{cb@lump.wustl.edu}
\affiliation{Department of Physics, Washington University, St. Louis, MO 63130, USA}
\author{C.~DeTar}
\affiliation{Department of Physics and Astronomy, University of Utah, Salt Lake City, UT 84112, USA}
\author{J.~Foley}
\affiliation{Department of Physics and Astronomy, University of Utah, Salt Lake City, UT 84112, USA}
\author{W.~Freeman}
\affiliation{Department of Physics, The George Washington University, Washington, DC 20052, USA}
\author{Steven Gottlieb}
\affiliation{Department of Physics, Indiana University, Bloomington, IN 47405, USA}
\author{U.M.~Heller}
\affiliation{American Physical Society, One Research Road, Ridge, NY 11961, USA}
\author{J.E.~Hetrick}
\affiliation{Physics Department, University of the Pacific, Stockton, CA 95211, USA}
\author{J.~Kim}
\affiliation{Department of Physics, University of Arizona, Tucson, AZ 85721, USA}
\author{J.~Laiho}
\affiliation{SUPA Department of Physics and Astronomy, University of Glasgow, Glasgow G12 8QQ, Scotland, UK}
\affiliation{Department of Physics, Syracuse University, Syracuse, NY 13244, USA}
\author{L.~Levkova}
\affiliation{Department of Physics and Astronomy, University of Utah, Salt Lake City, UT 84112, USA}
%DT-12/13/12 Ludmila is now paid by UA
\affiliation{Department of Physics, University of Arizona, Tucson, AZ 85721, USA}
\author{M.~Lightman}
\affiliation{Department of Physics, Washington University, St. Louis, MO 63130, USA}
\author{J.~Osborn}
\affiliation{Argonne Leadership Computing Facility, Argonne National Laboratory, Argonne, IL
60439, USA}
\author{S.~Qiu}
\affiliation{Department of Physics and Astronomy, University of Utah, Salt Lake City, UT 84112, USA}
\author{R.L.~Sugar}
\affiliation{Department of Physics, University of California, Santa Barbara, CA 93106, USA}
\author{D.~Toussaint}
\email{doug@physics.arizona.edu}
\affiliation{Department of Physics, University of Arizona, Tucson, AZ 85721, USA}
\author{R.S.~Van de Water}
\email{ruthv@fnal.gov}
\affiliation{Theoretical Physics Department, Fermi National Accelerator Laboratory, Batavia 60510, USA}
\author{R.~Zhou}
\affiliation{Department of Physics, Indiana University, Bloomington, IN 47405, USA}
\collaboration{MILC Collaboration}\noaffiliation

%\preprint{FERMILAB-PUB-12/333-T }

\title{\boldmath Leptonic decay-constant ratio $f_{K^+}/f_{\pi^+}$ from lattice QCD with physical light quarks}

\date{\today}

\begin{abstract} A calculation of the ratio of leptonic decay constants $f_{K^+}/f_{\pi^+}$ makes possible a
precise determination of the ratio of CKM matrix elements $|V_{us}|/|V_{ud}|$ in the Standard Model, and
places a stringent constraint on the scale of new physics that would lead to deviations from unitarity in the
first row of the CKM matrix.  We compute $f_{K^+}/f_{\pi^+}$ numerically in unquenched lattice QCD using
gauge-field ensembles recently generated that include four flavors of dynamical
quarks:  up, down, strange, and charm.  We analyze data at four lattice spacings $a \approx 0.06, 0.09,
0.12$, and 0.15~fm with simulated pion masses down to the physical value 135~MeV. We obtain $f_{K^+}/f_{\pi^+} = 1.1947(26)(37)$, where the errors are statistical and total systematic, respectively.  This is our first physics result from our $N_f = 2+1+1$ ensembles, and the first calculation of $f_{K^+}/f_{\pi^+}$ from lattice-QCD simulations at the physical point.  Our result
is the most precise lattice-QCD determination of $f_{K^+}/f_{\pi^+}$, with an error comparable to the current world average.  When combined with experimental measurements of the leptonic branching fractions,
it leads to a precise determination of $|V_{us}|/|V_{ud}| = 0.2309(9)(4)$ where the errors are
theoretical and experimental, respectively.
\end{abstract}

% insert suggested PACS numbers in braces on next line
\pacs{13.20.-v,	%Leptonic, semileptonic, and radiative decays of mesons
12.38.Gc,    %Lattice QCD calculations
12.15.Hh}	%Determination of Cabibbo-Kobayashi & Maskawa (CKM) matrix elements

% insert suggested keywords - APS authors don't need to do this
%\keywords{}

%\maketitle must follow title, authors, abstract, \pacs, and \keywords
\maketitle

%%%%%%%%%%%%
%%Motivation
%%%%%%%%%%%%

{\it Motivation}. ---  Leptonic decays of charged pseudoscalar mesons are sensitive probes of quark flavor-changing
interactions.   Experimental measurements of the leptonic decay widths, when combined with precise theoretical calculations
of the leptonic decay constants, enable the determination of elements of the Cabibbo-Kobayashi-Maskawa (CKM) quark-mixing
matrix.   Further, when combined with independent determinations of the CKM matrix elements from other processes such as
semileptonic meson decays, they make possible precise tests of the Standard-Model CKM framework.
Here we present a lattice-QCD calculation of the decay-constant ratio $f_{K^+}/f_{\pi^+}$, which may be used to determine $|V_{us}|/|V_{ud}|$ via~\cite{Marciano:2004uf,Aubin:2004fs}
\begin{equation}
\frac{\Gamma(K\to l \bar\nu_l)}{\Gamma(\pi\to l \bar\nu_l)} = \frac{|V_{us}|^2}{|V_{ud}|^2} \frac{f_K^2}{f_\pi^2} \frac{m_K}{m_\pi} \frac{\left( 1-\frac{m_l^2}{m_K^2} \right)^2}{ \left( 1-\frac{m_l^2}{m_\pi^2} \right)^2} \left[ 1 + \delta_{\rm EM} \right] \,,  \label{eq:Vus_Ratio}
\end{equation}
where $l = e,\mu$ and $\delta_{\rm EM}$ is a known 1-loop QED correction~\cite{Decker:1994ea,Finkemeier:1995gi}.  The
ratio $f_{K^+}/f_{\pi^+}$ can be calculated to subpercent precision using numerical lattice-QCD simulations~\cite{Laiho:2009eu,Colangelo:2010et} because the
Monte Carlo statistical errors are correlated between the numerator and denominator and the explicit dependence on the lattice
scale drops out.

Lattice QCD enables {\it ab initio} calculations of nonperturbative hadronic matrix elements.
The QCD Lagrangian is formulated in discrete Euclidean spacetime, after which the now finite-dimensional path integral can
be solved numerically with Monte Carlo methods.  Once the $N_f + 1$ parameters of the QCD action (quark masses and gauge
coupling) are fixed by matching to an equal number of experimental measurements, all other outputs of the lattice
simulation are predictions of the theory.  In practice, currently available computing resources limit the lattice spacing
in most recent simulations to $a \gtapprox 0.06$~fm, the spacetime volume of the lattice to $L^3 \sim (4~{\rm fm})^3$, and
the pion mass to $m_\pi \gtapprox 200$~MeV.  These choices introduce systematic errors that must be quantified, but the
good agreement between lattice-QCD calculations and experiment for a wide range of observables~\cite{Davies:2003ik,Kronfeld:2012uk}
demonstrates that the errors are controlled.

Many precise lattice-QCD weak-matrix element calculations relevant for quark-flavor physics have been obtained using our
$N_f = 2+1$ flavor gauge-field ensembles generated using the asqtad-improved staggered quark
action for the dynamical $u$, $d$, and $s$ quarks~\cite{Bernard:2001av,Aubin:2004wf,Bazavov:2009bb}. Our calculations of $f_{K+}/f_{\pi^+}$ on those ensembles \cite{Aubin:2004fs,Bazavov:2009bb,Bazavov:2010hj} ultimately reached a precision of $\sim\!0.6\%$.  We are now embarking on a new program of configuration generation using the highly-improved staggered quark (HISQ)
action~\cite{Follana:2006rc}, which has two primary advantages.  The most important staggered discretization errors are approximately  3 times smaller for the HISQ action than for the asqtad action at the same lattice spacing~\cite{Bazavov:2010ru}; this makes
it possible to undertake simulations at the physical pion mass, which require a box of approximately $(5~{\rm fm})^3$.  The improved
quark dispersion relation for the HISQ action also allows the inclusion of dynamical charm quarks in the simulations.   

One of the largest sources of uncertainty in lattice-QCD determinations of 
quantities such as $f_{K^+}/f_{\pi^+}$ 
is from the extrapolation of the
simulation results to the physical up- and down-quark masses. 
In this work, we replace this error with a  
negligible
interpolation error by simulating at light-quark masses very close to or even below their physical values. 
The calculation presented here, using the configurations with
  four flavors of dynamical HISQ quarks, lays the foundation for a larger quark
  flavor-physics program to compute pion, kaon, and charmed weak matrix elements,
  and eventually even those with bottom quarks, on these ensembles.

%%%%%%%%%%%%
%%Lattice calculation
%%%%%%%%%%%%

{\it Lattice-QCD Calculation}. --- This calculation is based on a subset of our (2+1+1)-flavor ensembles described in Refs.~\cite{Bazavov:2010ru,Bazavov:2012uw} .
These ensembles use a one-loop Symanzik-improved gauge action for the
gluons~\cite{Luscher:1985zq,Hart:2008sq}, and the HISQ action for the dynamical $u$, $d$, $s$, and $c$
quarks.  The simulated strange and charm sea-quark masses ($m_s$ and $m_c$) are fixed at approximately their
physical values.  The simulated up and down sea-quark masses are taken to be degenerate with a common mass $m_l \approx m_s/27$, such that the simulated pion mass is approximately the physical value
135~MeV~\cite{Beringer:1900zz}.  Additional ensembles with slightly heavier pions ($m_l = m_s/10$) allow us
to correct $f_{K^+}/f_{\pi^+}$ {\it a posteriori} for the slight mistuning of the simulated sea-quark
masses.  The spatial volumes are sufficiently large ($m_\pi L > 3.5$) that finite-volume corrections are
expected to be at the subpercent level for light pseudoscalar meson masses and decay constants, and we
confirm this with an explicit finite-volume study.  We use four lattice spacings $a\approx 0.15$, 0.12, 0.09, and 0.06~fm to enable a controlled extrapolation to
the continuum ($a \to 0$) limit.  The specific numerical simulation parameters are given in
Table~\ref{tab:LatParams}.

\begin{table} \caption{$N_f = 2 + 1 + 1$ gauge-field ensembles used in this work.  The lattice spacing in fm is
approximate.   Ensembles for which the number of equilibrated lattices ${N}_{\rm lats}$ has
reached 1000 are considered complete.  At $a \approx 0.12$~fm, we have three ensembles with identical simulation
parameters except for the spatial volume.  For these ensembles, the given pion masses and ${N}_{\rm lats}$ are approximate, since they differ slightly for the three volumes.
}
%\textcolor{magenta}{DT: The numbers in the HISQ config paper are probably Fp4s or r1 scale, numbers in
%lattice slides were Fk scale, to be consistent I give them here with Fpi scale.  EXCEPT, for the finite size
%test cases, I used the lattice spacing from the 32\^3 ensemble to convert all the pion masses to MeV so that
%the differences would just reflect the effect of the lattice size. RMS masses were determined from
%measurements on .4 m\_s ensembles, assuming differences independent of quark mass}
% DT 1/17/13 for the .06 fm physical ensemble, results on that ensemble were used for RMS masses
    \begin{tabular}{lcccr}
    \hline\hline
 	$a$(fm) & $L$(fm)& $M_{\pi, 5}$(MeV) & $M_{\pi, {\rm RMS}}$(MeV) & ${N}_{\rm lats}$ \\ % ens a_fm(Fpi) error
	\hline
	0.15 & 3.66 & 214 & 352 & 1000 \\	% l2448f211b580m0064m0640m828 0.15261 0.00020
	0.15 & 4.82 & 133 & 311 & 1000 \\\hline	% l3248f211b580m00235m0647m831 0.15078 0.00018
	0.12 & \{2.84,3.79,4.73\} & 215 & 294 & 1000 \\	
	0.12 & 5.83 & 133 & 241 & 1000 \\\hline	% l4864f211b600m00184m0507m628 0.12141 0.00012
	0.09 & 4.33 & 215 & 244 & 1000 \\	% l4896f211b630m00363m0363m430 0.09030 0.00018
	0.09 & 5.63 & 128 & 173 & 707  \\\hline	% l6496f211b630m0012m0363m432 0.08792 0.00010
	0.06 & 3.79 & 223 & 229 & 678  \\	% l64144f211b672m0024m024m286 0.05922 0.00012
	0.06 & 5.45 & 134 & 143 & 310  \\	% l96192f211b672m0008m022m260 0.05688 0.00017
    \hline\hline
    \end{tabular}
\label{tab:LatParams}
\end{table}

\begin{figure*}[t] \includegraphics[width=0.4\linewidth]{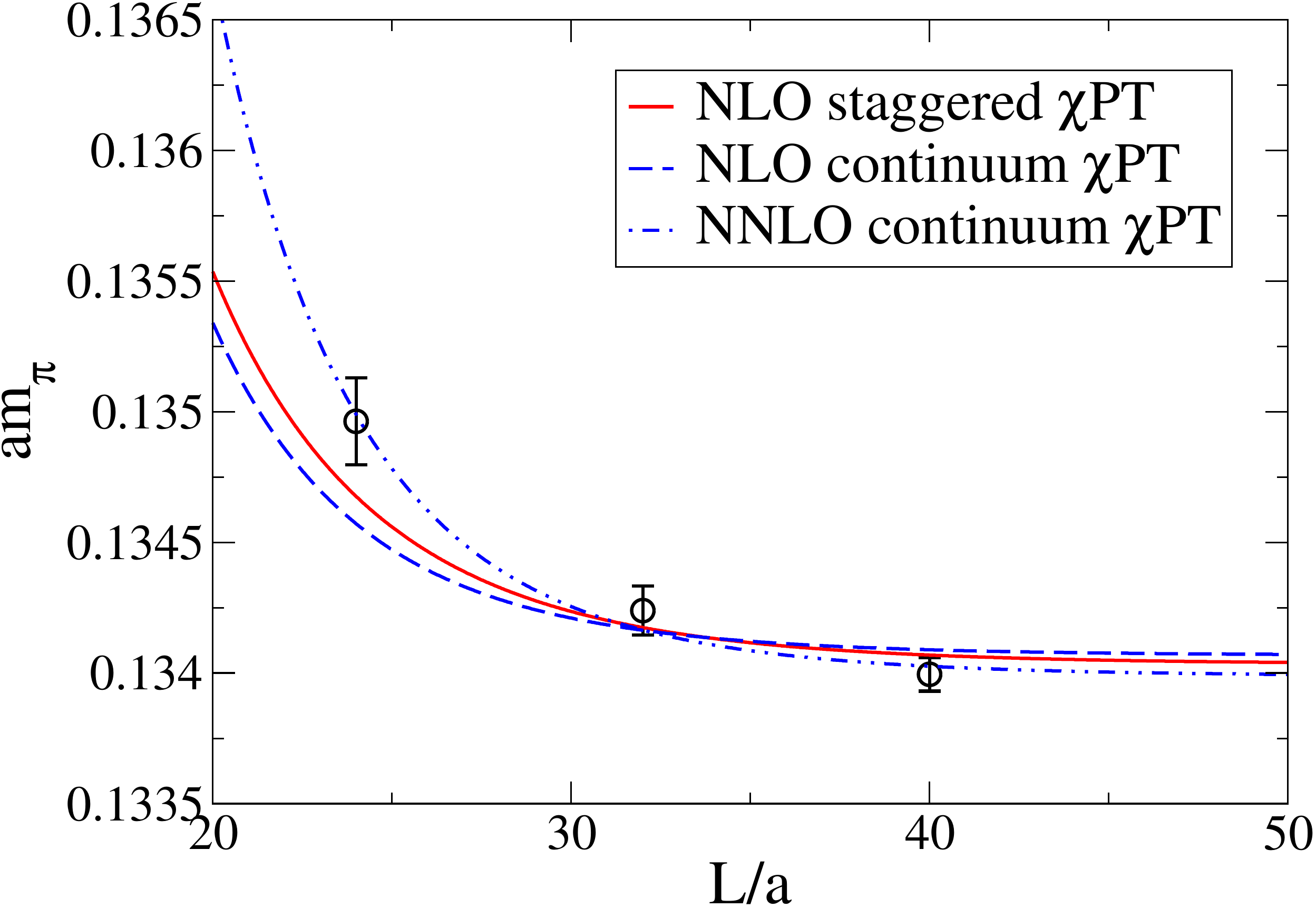}  $\qquad$
\includegraphics[width=0.4\linewidth]{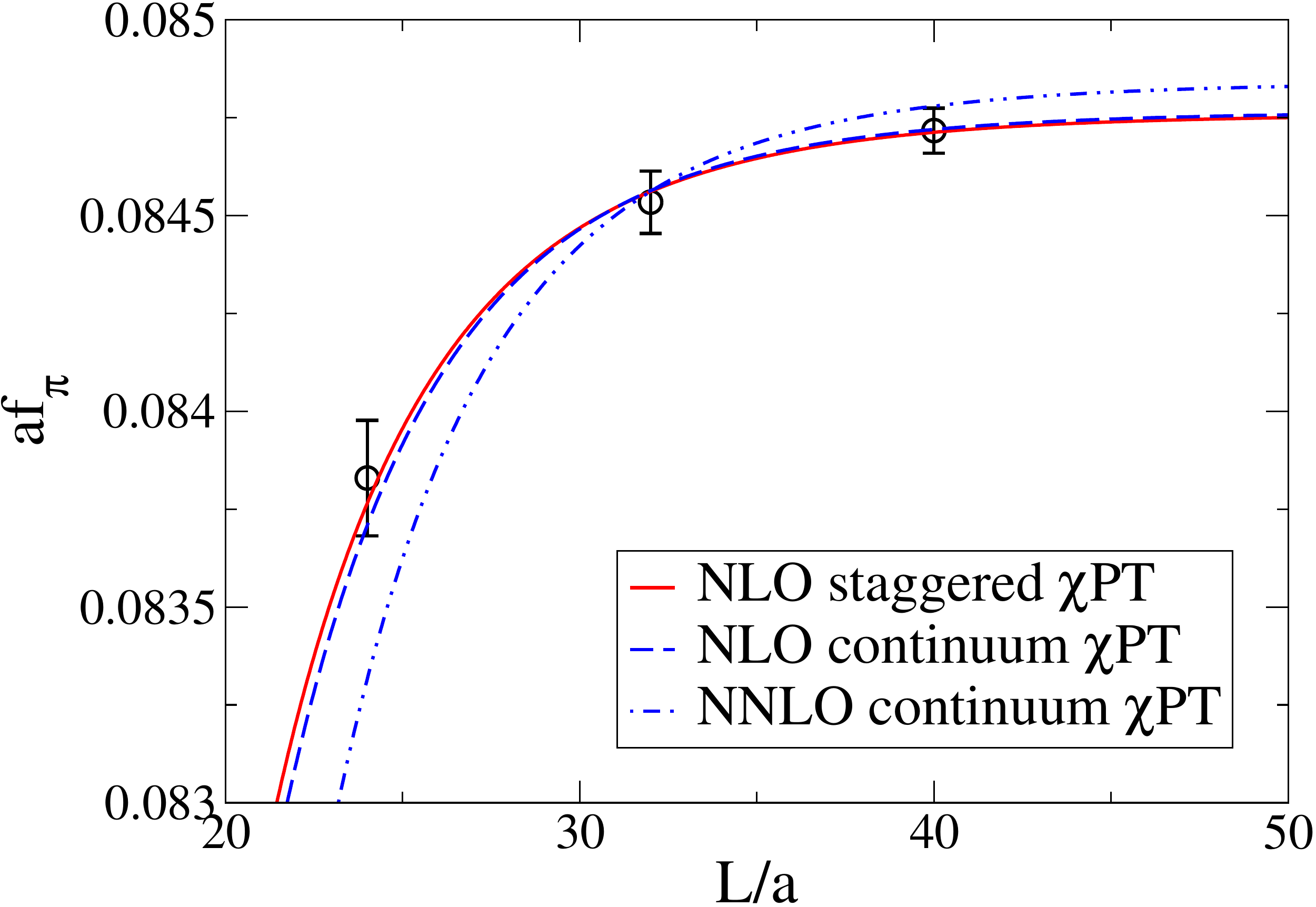} \caption{Pion mass (left) and decay constant (right)
versus lattice spatial extent.  The open circles show the numerical data with statistical errors.  The
staggered NLO $\chi$PT prediction is shown as a solid red line.  The continuum NLO (NNLO) $\chi$PT
predictions are shown as dashed (dot-dashed) blue lines.  The close agreement of the staggered
and continuum NLO $\chi$PT fits for $f_\pi$ is a numerical coincidence resulting from the cancellation of
effects from pion taste splittings and quark-disconnected hairpin contributions, and is not seen at other
lattice spacings.} \label{fig:FV_check} \end{figure*}

Naive lattice discretizations of the quark action suffer from the problem of fermion doubling.  The staggered quark action only partly eliminates fermion doublers, reducing the number of species from 16 to 4.  The remaining species are
referred to as ``tastes'' to distinguish them from physical flavors.  Quarks of different taste can interact
by exchanging gluons with momentum components close to the lattice cutoff $p_\mu \approx \pi/a$.
Taste-changing interactions split the mass degeneracy of pions
composed of different quark tastes.  The
taste splittings are of ${\mathcal O}(\alpha_S^2 a^2)$ for the HISQ action, and decrease rapidly with lattice
spacing.  The taste-Goldstone and root-mean-squared (RMS) sea-pion masses
are given in Table~\ref{tab:LatParams};  the difference between the two is only about 10~MeV for the finest $a \approx 0.06$~fm ensembles used in this analysis.   More details can be found in Ref.~\cite{Bazavov:2012uw}.

The dynamical HISQ simulations use the fourth-root procedure to remove the unwanted taste degrees of freedom.
Due to taste-changing interactions, the rooted Dirac operator is nonlocal at nonzero lattice spacing, and
leads to violations of unitarity~\cite{Prelovsek:2005rf,Bernard:2006zw,Bernard:2006ee,Bernard:2007qf}.
Nevertheless, there are strong theoretical arguments and numerical evidence that the continuum limit of the
rooted staggered lattice theory is indeed
QCD~\cite{Shamir:2004zc,Shamir:2006nj,Follana:2004sz,Durr:2004as,Durr:2004ta,Wong:2004nk,Durr:2006ze,Donald:2011if}.  

We obtain the meson masses and decay constants from fits of two-point correlation functions with a pseudoscalar interpolating operator at both the source and sink.  For each valence-quark mass combination, we
compute the decay
constant using a partially-conserved axial current relation:
\begin{equation}
	F_{xy} = (m_x + m_y) \langle 0 | \bar{x} \gamma_5 y | P_{x \bar{y}} \rangle / M_{xy}^2 \,, \label{eq:fPS}
\end{equation}
where $M_{xy}$ is the mass of the pseudoscalar meson $P_{x \bar{y}}$ with bare valence-quark masses $m_x$ and $m_y$.  We choose a fitting range for the correlators such that, in the case of degenerate valence-quark masses, a single-exponential fit 
gives a good fit as measured by the correlated  $\chi^2/{\rm dof}$ and $p$-value.
We include both the ground state and an opposite-parity excited state for the nondegenerate correlators.  
Statistical errors are estimated by jackknifing.
The ground-state mass and amplitudes are stable with respect to a reasonable reduction of the minimum time separation of the source and sink, and with the inclusion of additional excited states;  therefore the error due to excited-state
contamination can be neglected compared with the statistical errors.  
Additional details of these fits can be found in Ref.~\cite{Bazavov:2012us}.

Chiral perturbation theory ($\chi$PT) predicts that the finite-volume corrections to pion and kaon masses and decay
constants are at the per mill level in our simulations, and are therefore comparable to the size of the statistical errors
in our numerical data.  We check this expectation with an explicit comparison of these quantities measured on three $a
\approx 0.12$~fm ensembles with identical simulation parameters except for the spatial lattice volume (see
Table~\ref{tab:LatParams}).  We fit the data at the three volumes to the functional form:
\begin{equation}
	aX(L) \! = \! A_X\left(1 + B_X(m_\pi, L)\right) 
\end{equation}
where $X = \{m_\pi, f_\pi, m_K, f_K\}$, $A_X$ is a free parameter, and the 
function $B_X$ is determined at a given order in $\chi$PT.
We try both the NLO  expression for
$B_X$ in staggered $\chi$PT \cite{Aubin:2003mg,Aubin:2003uc}, and the continuum
NLO and resummed NNLO expressions~\cite{Colangelo:2005gd}.
Figure~\ref{fig:FV_check} compares the numerical lattice data for $m_\pi$ and $f_\pi$ with these expressions.  NLO
staggered $\chi$PT describes the $f_\pi$ data very well at all three volumes, and describes the $m_\pi$, $m_K$, and $f_K$
data adequately (to $\sim\!0.4\%$ or better).   We therefore use NLO staggered finite-volume $\chi$PT to correct our
simulation data in our central fit, and use the continuum NLO and NNLO finite-volume $\chi$PT corrections to estimate the
systematic uncertainty.   

Because our lattice-QCD simulations are isospin-symmetric
(the up and down sea-quark masses are equal), we
adjust the experimental inputs to what they would be in
a world without electromagnetism or isospin violation before matching the simulation data to experiment to
find the strange quark mass $m_s$ and the average light quark mass $\hat m= (m_u + m_d)/2$.  We follow the
approach of Ref.~\cite{Aubin:2004fs}, but with updated values for the EM correction to Dashen's theorem.
%\textcolor{blue}{Specifically, we do not adjust the neutral pion mass because the leading-order isospin correction to $M_{\pi^0}^2$ is $\propto (m_u - m_d)^2/\Lambda^2_\chi$ in $\chi$PT and therefore small, and the electromagnetic corrections vanish in the chiral limit for neutral mesons and are thus also small.}  
For the kaon, we consider the isospin-averaged mass $M_{\widehat{K}}^2 = (M_{K^+}^2 + M_{K^0}^2)_{\rm QCD}/2$, where the subscript ``QCD" indicates that the leading EM effects in the masses were removed
from the experimental masses \cite{Beringer:1900zz};  we take the
parameter $\Delta_{\rm EM} = 0.65(7)_{\rm stat}(14)_{\rm syst}$ that characterizes violations of Dashen's
theorem from our ongoing lattice QED+QCD simulations using asqtad sea quarks~\cite{Basak:2012zx}.

\begin{figure} \includegraphics[width=\linewidth]{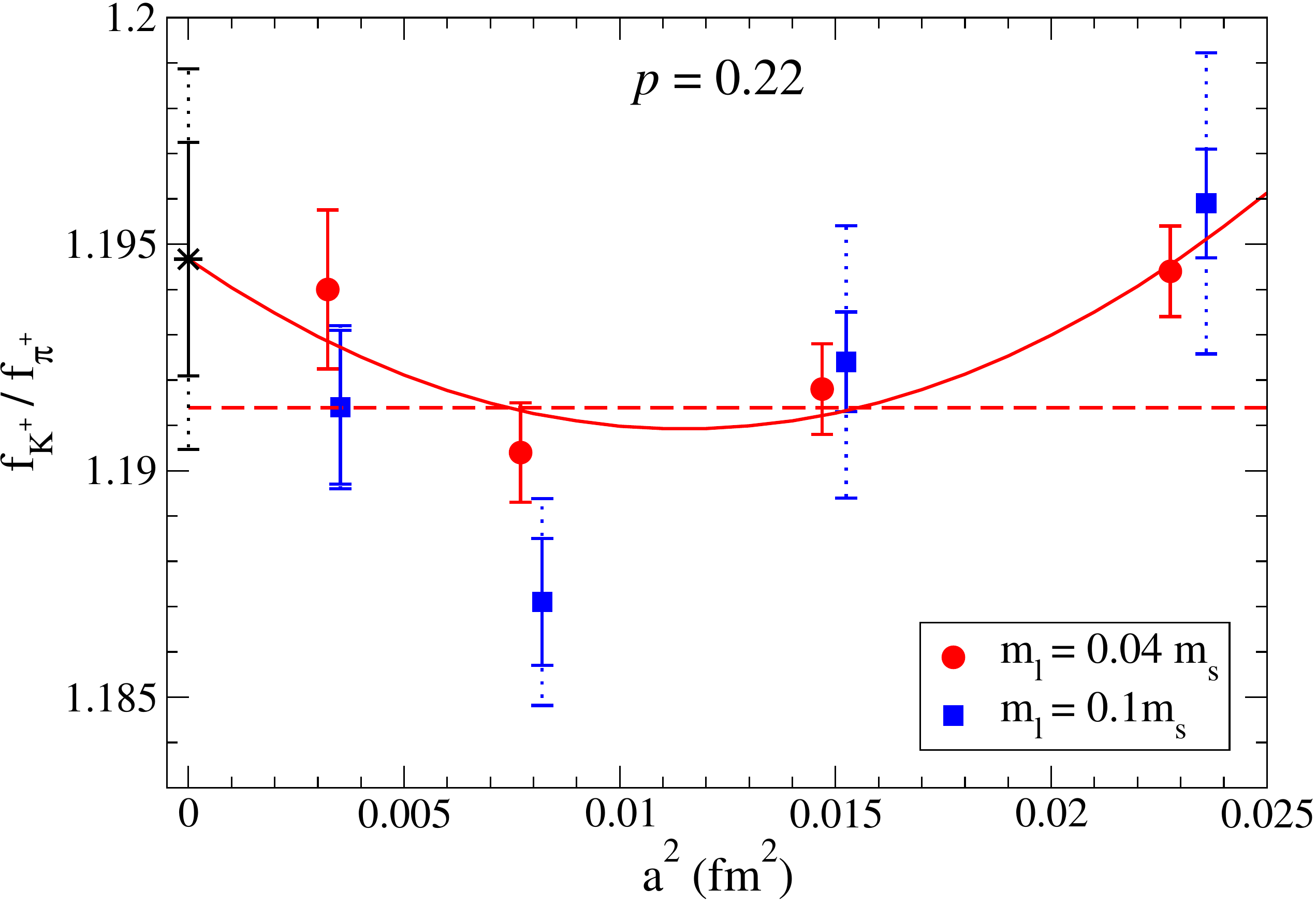} \caption{Continuum extrapolation of
$f_{K^+}/f_{\pi^+}$.  Our central value is obtained from a quadratic-in-$a^2$ fit to the physical-quark-mass data
points (red circles), adjusted for tuning errors to the physical sea quark masses using the $0.1m_s$ data points (blue
squares). The size of the adjustments is too small to be visible on this plot.   The black star shows
$f_{K^+}/f_{\pi^+}$ at the physical quark masses in the continuum, where the inner solid error bar is statistical and
the outer dotted error bar includes the continuum-extrapolation error added in quadrature.  An alternative,
constant-in-$a^2$ fit to the finest two physical-mass data points, is shown by the dashed horizontal line.  Its
deviation from the central value gives our estimate of the continuum extrapolation error (see text).
The dotted error bars on the $0.1m_s$ data show the systematic error from the valence-quark mass tuning  added in quadrature to the statistical error (see text); the tuning uncertainty on the physical-mass ensembles is negligible.
}
\label{fig:a2_extrap} \end{figure}

In this work, we tune the quark masses and lattice scale and determine $f_{K^+}/f_{\pi^+}$ in a single,
self-contained analysis as follows.  We begin with numerical lattice data for meson masses and decay constants for a selection of valence-quark masses that allow us to adjust for slight mistunings and interpolate or extrapolate to the physical values.  In order to reduce autocorrelations below measurable
levels, the single-elimination jackknife results are blocked by 20 lattices (100 or 120 molecular dynamics
time units) before the final averaging.  The physical-quark-mass $a\approx0.06$ fm ensemble has an insufficient number of configurations for blocks that large. Instead, we block by 5 trajectories on that ensemble and rescale the resulting errors by a factor of 1.17, which is determined by comparing the errors at block sizes around 20 to those at block size 5 on the $m_l=0.1m_s$, $a\approx0.06$ fm ensemble, as well as an additional $m_l=0.2m_s$, $a\approx0.06$ fm ensemble not otherwise discussed here.
%DT fixed here for "lin0" tuning
We then construct the squared ratio of pseudoscalar-meson
mass to decay constant, $M_{xx}^2 / F_{xx}^2$, which $\chi$PT predicts to be approximately
proportional to the valence-quark mass $m_x$.  Using the lightest valence-quark mass on each ensemble, we interpolate or extrapolate data for this ratio linearly in $m_x$ to where it equals the experimental value for $M_{\pi^0}^2/f_{\pi^+}^2$.   This fixes the physical value of $\hat{m}$.  We obtain the lattice spacing from requiring that the decay constant at this mass equal the experimental value for $f_{\pi^+}$~\cite{Rosner:2010ak}, where we use a quadratic interpolation/extrapolation through the lightest three valence-quark masses.  We then take nondegenerate ``kaons" in which the strange
valence-quark mass ($m_y$) is 1.0 or 0.8 times the strange sea-quark mass and the lighter valence-quark mass ($m_x$) is the lightest available, and linearly interpolate in the valence strange-quark mass to where $2M_{xy}^2 - M_{xx}^2 = 2M_{\widehat{K}}^2 - M_{\pi^0}^2$;  this fixes the physical strange-quark mass $m_s$.  Once we know $m_s$, we obtain the $u$-$d$ mass-splitting from the difference between the EM-subtracted neutral and charged kaon masses: $m_d - m_u = (M_{K^0}^2 - M_{\hat K}^2)_{\rm QCD} /  (\partial M^2_{sx} / \partial m_x)$.  Finally, we obtain $f_{K^+}$ by linearly interpolating the decay constant $F_{xy}$ in the light valence-quark mass to $m_u$ and in the strange valence-quark mass to $m_s$.  In $f_{K^+}$, we neglect isospin violations from the sea quarks, and we neglect all isospin violations in $f_{\pi^+}$, because these effects are of NNLO in $\chi$PT~\cite{Colangelo:2010et}.  We also
slightly correct $f_K$ {\it a posteriori} for sea-quark mass mistuning using the slope with respect to the light
sea-quark mass, $\partial f_K/ \partial m_l$, measured by comparing the value of $f_K$ from the physical-mass ensembles with that from ensembles with $m_l \approx 0.1m_s$.  The results for $f_{K^+}/f_{\pi^+}$ are shown versus $a^2$ in Fig.~\ref{fig:a2_extrap}.  

We extrapolate $f_{K^+}/f_{\pi^+}$ at physical quark masses to the continuum limit quadratically in $a^2$.  We
estimate the uncertainty in the continuum extrapolation from alternative fits with different ans{\"a}tze for the $a^2$
dependence and excluding the coarsest ensemble(s).  Two reasonable choices are a linear extrapolation of the three
physical-mass data points with $a\le0.12$ fm, and a constant fit to the two physical-mass data points with $a\le 0.09$
fm. Of these, the constant fit, shown in Fig.~\ref{fig:a2_extrap}, gives the larger difference from the central
extrapolation, $0.28\%$.  Additional extrapolations have  also been tried, including a quadratic-in-$a^2$, linear in
$m_l$ fit to all the $m_l\approx m_s/27$ and $m_l=0.1m_s$ data points, and a similar fit restricted to ensembles with
$a\le0.12$ fm.  These fits give differences from the central value that are comparable to or smaller than the difference seen
above, as does a simple linear-in-$a^2$ extrapolation of the two finest $a \leq 0.09$ fm physical-mass points.  We
thus take $0.28\%$ as our estimate of the discretization error.  Note that, when the scale-setting procedure is changed by repeating the analysis using
other quantities to fix the lattice scale instead of $f_\pi$ (see Ref.~\cite{Bazavov:2012uw} for further details), we
find a $0.17\%$ difference with the central value, well within the estimate for the continuum extrapolation error.  To
estimate the finite-volume error, we repeat the entire analysis using finite-volume corrections for the pion (kaon)
masses and decay constants calculated at NNLO (NLO) in continuum $\chi$PT.   We propagate the statistical
uncertainties in the tuned quark masses and lattice scale throughout the analysis via the jackknife method, so they
are already folded into the statistical error.
We estimate the systematic uncertainty from the quark-mass tuning and scale setting by taking the difference of our central
tuning procedure with one where we interpolate/extrapolate the ratio $M^2_{xx}/F^2_{xx}$ quadratically in $m_x$.  For the physical quark-mass ensembles (and hence for the continuum extrapolated result), the valence-quark masses are very close to the tuned physical values, so the choice of interpolation fit function makes a negligible difference.  On the 0.1 $m_s$ ensembles, however, valence masses as light as physical are not available, so the choice of fit function for the valence-quark extrapolation has a greater impact.  The systematic errors from the tuning procedure are plotted as dotted error bars in Fig.~\ref{fig:a2_extrap}.
We estimate the uncertainty due to EM effects by
varying the values of the EM-subtracted meson masses used in the quark-mass tuning; this primarily affects $m_u$.  We
vary the parameter $\Delta_{EM}$ by its total error~\cite{Basak:2012zx}.  We also repeat the tuning including a
previously-neglected EM correction to the neutral kaon mass, $\delta M^2_{K^0, {\rm EM}}=901(8)_{\rm stat}~{\rm
MeV}^2$~\cite{Bernard:CD12}.  Note that direct EM effects on the weak matrix elements are by definition removed from
$f_{\pi^+}$ and $f_{K^+}$ \cite{Beringer:1900zz}. 

%%%%%%%%%%%%
%%Results and conclusions
%%%%%%%%%%%%

{\it Results and Conclusions}. ---
We obtain
the following determination of the ratio $f_{K^+}/f_{\pi^+}$:
\begin{equation}
	f_{K^+}/f_{\pi^+} = 1.1947(26)_{\rm stat} (33)_{a^2} (17)_{\rm FV}
         (2)_{\rm EM} , \label{eq:result}
\end{equation}
with an approximately 0.4\% total uncertainty.  Our result agrees with, but is more precise than, other independent
lattice-QCD calculations~\cite{Follana:2007uv,Durr:2010hr,Aoki:2010dy,Bazavov:2010hj,Laiho:2011np}, and its error is
competitive with that of the current lattice-QCD world average~\cite{Laiho:2009eu,Colangelo:2010et}. 
Because we use
$f_\pi$ to set the lattice scale, we also obtain a result for 
$f_{K^+} = 155.80(34)_{\rm stat}(48)_{\mathrm{sys}}(24)_{f_\pi}$
where the errors are due to statistics, systematics in $f_K/f_\pi$, and the uncertainty in $f_\pi$.  
This agrees with the experimental determination of $f_{K^+}$ assuming CKM unitarity~\cite{Rosner:2010ak}.

By combining $f_{K^+}/f_{\pi^+}$ from Eq.~(\ref{eq:result}) 
with recent experimental results for the leptonic 
branching fractions~\cite{Antonelli:2010yf}, 
we obtain $|V_{us}|/|V_{ud}| = 0.2309(9)_{\rm theo.} (4)_{\rm exp.}$.  
Taking $|V_{ud}|$ from nuclear $\beta$ decay~\cite{Hardy:2008gy},
 we also obtain $|V_{us}| = 0.2249(8)_{\rm theo.} (4)_{\rm exp.} (1)_{V_{ud}}$,
which 
agrees with and
has comparable errors to the determination from $K\to\pi\ell\nu$ semileptonic
decay~\cite{Lubicz:2009ht,Boyle:2010bh,Antonelli:2010yf,Bazavov:2012cd}.
Further, our result places stringent constraints on new-physics scenarios that would lead
to deviations from unitarity in the first row of the CKM matrix. 
We find $1 - |V_{ud}|^2 - |V_{us}|^2 - |V_{ub}|^2 =  0.0003(6)$, which is in excellent agreement with the Standard-Model value of zero.
\\ \indent We are currently extending the ensemble with the physical pion mass at $a \approx 0.06$~fm as well as the other ensembles with fewer than 1000 equilibrated lattices. We will include this additional data in a longer work that updates this result and presents determinations of the charmed pseudoscalar decay constants.  We will also perform a more sophisticated analysis using staggered chiral perturbation
theory~\cite{Aubin:2003mg,Aubin:2003uc,Komijani:2012fq} to obtain the low-energy constants of $\chi$PT.  
Our result for $f_{K^+}/f_{\pi^+}$ lays the foundation for a new large-scale lattice-QCD physics program using the $N_f = 2+1+1$ flavor HISQ
gauge-field ensembles with physical pion masses.   These ensembles will enable calculations of light and heavy-light
hadronic weak matrix elements with unprecedented precision, and will ultimately strengthen tests of the Standard Model and
improve the reach of searches for new physics in the quark-flavor sector.  

NB: After this work was submitted for peer review, another determination of $f_K/f_\pi$ using our HISQ ensembles and consistent with our result was posted by HPQCD~\cite{Dowdall:2013rya}.

%%%%%%%%%%%%
%%Acknowledgements
%%%%%%%%%%%%

{\it Acknowledgments}. --- We thank Christine Davies and Andreas Kronfeld for useful discussions and comments on the manuscript.  We thank Maarten Golterman for pointing out a critical typo in the abstract.
Computations for this work were carried out with resources provided by 
the USQCD Collaboration, the Argonne Leadership Computing Facility and
the National Energy Research Scientific Computing Center, which
are funded by the Office of Science of the United States Department of Energy; and with resources
provided by
the National Center for Atmospheric Research,
the National Center for Supercomputing Applications,
the National Institute for Computational Science,
and the Texas Advanced Computing Center, 
which are funded through the National Science Foundation's Teragrid/XSEDE and Blue Waters Programs. 
We thank the staffs of NICS, ALCF and NCSA for their assistance with block
time grants and Early Science usage.
This work was supported in part by the U.S. Department of Energy under Grants 
No.~DE-FG02-91ER40628 (C.B., M.L.), 
No.~DE-FC02-06ER41446 (C.D., L.L., J.F.), 
No.~DE-FG02-91ER40661 (S.G., R.Z.), 
No.~DE-FG02-85ER40237 (J.L.),
%DT-12/13/12 Scidac grant for Alexei, Jongjeong and Ludmila
No.~DE-FG02-04ER-41298 (D.T.); 
No.~DE-FC02-06ER-41439 (J.K.,A.B.,L.L),
by the National Science Foundation under Grants 
No.~PHY-1067881, No.~PHY-0757333, No.~PHY-0703296 (C.D., L.L., J.F., S.Q.), 
%DT-12/13/12 NSF PIF for Alexei and Jongjeong
No.~PHY-0555397, (A.B.),
No.~PHY-0903536, (A.B.,J.K.),
No.~PHY-0757035 (R.S.),
and by the Science and Technology Facilities Council and the Scottish Universities Physics Alliance (J.L.).
% Brookhaven boilerplate for Alexei
This manuscript has been co-authored by employees of Brookhaven Science Associates, LLC,
under Contract No.~DE-AC02-98CH10886 with the U.S. Department of Energy. 
% Fermilab boilerplate for Ruth
Fermilab is operated by Fermi Research Alliance, LLC, under Contract No.~DE-AC02-07CH11359 with
the U.S. Department of Energy.

\bibliography{HISQ_fPS}
\bibliographystyle{apsrev4-1} % bst file

\end{document}